\documentclass{aa}
\usepackage{epsf}

\newcommand{\Nh}{\mbox{$N_{\rm H}$}}
\newcommand{\La}{\mbox{${\rm Ly\alpha}$}}
\newcommand{\ecsa}{\mbox{$\rm ergs\;cm^{-2}s^{-1}\mbox{\AA}^{-1}$}}
\newcommand{\ecs}{\mbox{$\rm ergs\;cm^{-2}s^{-1}$}}
\newcommand{\es}{\mbox{$\rm ergs\;s^{-1}$}}

\newcommand{\kms}{\mbox{$\rm km\,s^{-1}$}}
\begin{document}

\thesaurus{08      
          (02.01.2 
           08.02.1 
           08.09.2 
           13.21.5 
           13.25.5 
          )}
\title{Time-resolved HST and IUE UV spectroscopy of the Intermediate Polar FO\,
Aqr\thanks{Based
on observations collected with the NASA/ESA Hubble Space Telescope, obtained
at the Space Telescope Science Institute and with the International Ultraviolet
Explorer, obtained from the IUE Final Archive at VILSPA}}

\author{
D. de Martino\inst{1},
R. Silvotti\inst{1},
D.A.H Buckley\inst{2},
B.T. G\"ansicke\inst{3}
M. Mouchet\inst{4},
K. Mukai\inst{5,6},
S.R. Rosen\inst{7}
}

\offprints{D.~de Martino}

\institute{
Osservatorio Astronomico di Capodimonte, Via Moiariello 16, I-80131 Napoli, Italy
   \and
South African Astronomical Observatory, PO Box 9, Observatory 7935, Cape Town,
South Africa
   \and
Universit\"ats Sternwarte G\"ottingen, Geismarlandstr. 11, D-37083, G\"ottingen,
Germany
   \and
DAEC, Observatoire de Paris, Section de Meudon, F-92195 Meudon Cedex,
France
   \and
Laboratory for High Energy Physics, NASA/GSFC, Code 662, Greenbelt, MD 20771,
USA
   \and
 Universities Space Research Association 
\and
Department of Physics and Astronomy, University of Leicester, University
Road, Leicester LE1 7RH, UK 
}

\date{Received June 15, 1999 ; accepted September 2, 1999 }

\authorrunning{de Martino et al.}
\titlerunning{Time Resolved HST and IUE ...}
\maketitle 

\begin{abstract}

Time resolved spectroscopy of the Intermediate Polar FO\,Aqr reveals
the presence of multiple periodicities in the UV range. A strong
orbital modulation dominates both continuum and emission line flux
variabilities, while line velocity motions are only detected at the
rotational frequency. A prominent orbital periodicity is also observed
in coordinated optical photometry, where FO\,Aqr was previously found
to be spin dominated.  The spectral dependence of the main
periodicities shows the presence of multi-temperature components in
FO\,Aqr and for the first time a hot and a cool component in the
rotational modulation.  From a comparison with previous UV and optical
data obtained in 1990, no spectral variations in the orbital and
rotational variabilities are detected, indicating no significant
changes in the effects of X-ray illumination but rather a shrinking of
the accretion curtain accompained by an increase in size of the
thickened part of the accretion disc.  These observations, consistent
with the recently discovered long term trend in the X-ray pulsation
amplitudes, independently confirm a change in the accretion mode in
FO\,Aqr, which switched from a disc-fed into a disc-overflow state,
likely triggered by mass accretion variations.

\end{abstract}

\section{Introduction}

Intermediate Polars (IPs) are a subclass of magnetic Cataclysmic 
Variables which consist
of an asynchronously rotating  ($\rm P_{spin} < P_{orb}$) 
magnetized white dwarf accreting from 
a late type, main sequence, Roche-lobe filling secondary star
(Patterson 1994; Warner 1995).

Except for a few systems
for which polarized optical/IR emission is detected, 
the white dwarf is believed to possess a 
weak ($\la$ 2\,MG) magnetic field which dominates the accretion flow 
only at a few  radii from its surface. Within the magnetospheric
radius, material is channeled towards the magnetic polar regions in an 
arc-shaped accretion curtain (Rosen et al. 1988). At
larger distances, different accretion patterns can be present:
 a truncated accretion disc (disc-fed systems), direct 
accretion from the stream onto the magnetosphere (disc-less systems) as
well as a combination of the two, where the stream material overpasses the
disc (disc-overflow) (Hellier 1995 and references therein). 

Due to the asynchronous rotation, IPs show a wide range of periodicities
at the white dwarf spin ($\omega$), the orbital
($\Omega$) and sideband frequencies (Warner 1986; Patterson 1994, Warner 
1995), whose
amplitudes can be different in different spectral ranges (de Martino 1993).
FO\,Aqr (H2215-086) was known to show strong periodic X-ray, optical and
IR  pulsations at the spin frequency, $\rm \omega = 1/(P_{spin}=20.9\,min)$, 
and lower amplitude variations at the orbital 
$\rm \Omega = 1/(P_{\rm orb}=4.85\,hr)$
and beat  $\rm \omega - \Omega = 1/(P_{beat}=22.5\,min)$ frequencies
(de Martino et al. 1994, hereafter Paper\,1, and references therein; 
Marsh \& Duck 1996; Patterson et al. 1998). The dominance of the spin pulsation at optical
and high X-ray energies ($>$5\,keV) can be accounted for by 
a disc-fed accretion, whose evidence was provided
by a partial eclipse  in the optical continuum and emission 
lines (Hellier et al. 1989; Mukai et al. 1994; Hellier 1995). 
FO\,Aqr was also found to possess a disc-overflow accretion mode 
(Hellier 1993), 
and the recent evidence  of 
a long term variability in the amplitudes of the X-ray pulsations 
has been interpreted as changes in the accretion mode (Beardmore et al. 1998). 
This kind of variability, only recently
recognized, is also observed in other IPs like  TX\,Col (Buckley 1996;
Norton et al. 1997) 
and BG\,CMi (de Martino et al. 1995). 

The identification of the actual accretion geometry and the determination
of  energy budgets of the primary X-ray and secondary reprocessed 
UV, optical and IR  emissions reside on multi-wavelength observations.
 The main modulations in FO\,Aqr have been 
studied in the X-rays and at optical/IR wavelengths.
The optical spin pulsations, occurring mostly in phase
with the X-ray ones, were found to arise from the outer regions of
the accretion curtain (Paper\,1; Welsh \& Martell 1996). 
The orbital modulation from UV to IR was instead found to be 
multicomponent and attributed 
to the X-ray heated azimuthal structure of the accretion disc (henceforth 
bulge) and to the inner illuminated face of the secondary
star (Paper\,1). However the temperature of the UV emitting
bulge could not be constrained due to the lack of a precise quantification of 
the spectral shape of the UV spin modulation.
In this work we present high temporal resolution spectroscopy
acquired with HST/FOS which provides the first detection of different
UV periodicities in FO\,Aqr. For a comprehensive study, 
these data are complemented with low temporal resolution IUE spectra 
along the orbital period. Coordinated optical photometry 
extends the study to a wider spectral range, providing the link to 
investigate the long term behaviour of these variabilities.

\section{Observations and data reduction}

The UV and optical campaign on FO\,Aqr was carried out 
between September and October 1995 with HST, IUE and at the
South African Astronomical Observatory. The journal of the observations is 
reported in Table\,1.

\begin{table}[h]     
\caption[]{Journal of observations.}
\begin{flushleft}
\begin{tabular}{lllcl}
\hline
\noalign{\smallskip}
Instrument  & Date  & UT$_{\rm start}$ & Duration \\
&       & hh:mm:ss    & s \\
\noalign{\smallskip}
\hline
\noalign{\smallskip}
{\em HST/FOS} & & & \\
\noalign{\smallskip}
Slot 1 & 1995 Sept. 10 & 14:11:04 & 2332 \\
Slot 2 & 1995 Sept. 10 & 15:45:11 & 2461 \\
Slot 3 & 1995 Sept. 10 & 17:21:18 & 2461 \\
Slot 4 & 1995 Sept. 10 & 18:57:49 & 2461 \\
Slot 5 & 1995 Sept. 10 & 20:34:22 & 2461 \\
Slot 6 & 1995 Sept. 10 & 22:10:54 & 2461 \\
\noalign{\smallskip}
\hline
\noalign{\smallskip}
{\em IUE} & & & \\
Image \# & & &\\
\noalign{\smallskip}
SWP56089 & 1995 Oct. 15 & 14:44:13 & 2520\\
SWP56090 & 1995 Oct. 15 & 16:21:26 & 2520\\
SWP56091 & 1995 Oct. 15 & 18:01:17 & 2520\\
SWP56092 & 1995 Oct. 15 & 19:22:39 & 2520\\
SWP56093 & 1995 Oct. 15 & 20:38:56 & 2520\\
SWP56094 & 1995 Oct. 15 & 22:02:27 & 2520\\
SWP56095 & 1995 Oct. 15 & 23:29:51 & 2520\\
SWP56096 & 1995 Oct. 16 & 00:57:16 & 2520\\
SWP56097 & 1995 Oct. 16 & 02:24:39 & 2520\\
SWP56098 & 1995 Oct. 16 & 03:55:35 & 2520\\
LWP31592 & 1995 Oct. 15 & 13:45:09 & 2520\\
LWP31593 & 1995 Oct. 15 & 15:34:55 & 2520\\
LWP31594 & 1995 Oct. 15 & 17:17:11 & 2520\\
LWP31595 & 1995 Oct. 15 & 18:45:29 & 1260\\
LWP31596 & 1995 Oct. 15 & 20:06:28 & 1260\\
LWP31597 & 1995 Oct. 15 & 21:25:29 & 1260\\
LWP31598 & 1995 Oct. 15 & 22:46:08 & 2520\\
LWP31599 & 1995 Oct. 16 & 00:13:32 & 2520\\
LWP31600 & 1995 Oct. 16 & 01:41:01 & 2520\\
LWP31601 & 1995 Oct. 16 & 03:26:04 & 1260\\
\noalign{\smallskip}
\hline
\noalign{\smallskip}
{\em SAAO} & & & \\
\noalign{\smallskip}
BVRI & 1995 Oct. 18 & 18:37:44 & 13758\\ 
     & 1995 Oct. 19 & 19:41:11 & 2745 \\
     & 1995 Oct. 20 & 20:01:30 & 7480 \\
No Filter & 1995 Oct. 21 & 18:04:35 & 15585\\
          & 1995 Oct. 23 & 18:40:20 & 9130\\

\noalign{\smallskip}
\hline
\end{tabular}
\end{flushleft}
\end{table}

\subsection{ The HST data}

HST Faint Object Spectrograph observations of FO\,Aqr were performed
on September 10, 1995. The observations were carried out in the rapid 
mode during 7 consecutive HST orbits. Due to target acquisition procedures 
the total effective on source exposure time was 4.07\,h, yielding 
six continuous exposure slots as
detailed in Table\,1. The orbital period was unevenly sampled since 
it is commensurable with the HST orbit. 
The G160L grating  was operated with the blue digicon covering the
range 1154--2508 \AA\ at a resolution of 6.8 \AA\ diode$^{-1}$ and with
the 0.86" upper square aperture, supposed to be free from the 
 1500--1560\, \AA \, photocathode blemish which is known to affect 
 the circular apertures (HST Data Handbook, 1997). A total of 797 spectra 
were collected, each  with an effective exposure time of $\sim$ 18\,s. 

The standard routine processing ST\,ScI pipeline applied to the data 
at the time of the observations revealed the presence of anomalous features 
when comparing the reduced spectra with the IUE data, and in 
particular in the regions 1500--1590\,\AA\ and 1950--2010\,\AA\ .  The data 
were  then re-processed  using the STSDAS/CALFOS routine within IRAF 
using the  latest reference files for sensitivity correction and appropriate
aperture  flatfield provided by the ST\,ScI Spectrograph Group in summer 1998. 
The calibrated 
spectra appear then free from the above features although the IUE fluxes are on 
average $\sim 1.1$ times larger than the FOS ones (see Fig.\,1). 
A check against systematic effects using spectra 
of standard stars observed with both IUE and FOS only 
confirms the known lower flux (on average of $\sim 7\%$) of IUE, 
with respect to that of FOS (Gonzalez-Riestra 1998). 
The residual flux difference may be due to the fact that the 
IUE data were acquired 
a month later than the FOS spectra.

The G160L grating provides zero order light covering the full range
between 1150 and 5500\,\AA, with an effective wavelength at 3400\,\AA,
which provides useful simultaneous broad-band photometry.
The flux calibration of the zero-order signal 
(Eracleous \& Horne 1994) updated for errors and post-COSTAR 
sensitivity and aperture throughput (HST Data Handbook 1997) has 
been applied to the signal extracted in the zero-order feature
of each of the 797 exposures.

\subsection{ The IUE data}

\noindent On October 15, 1995 ten IUE SWP (1150--1980\,\AA) and ten LWP
(1950--3200\,\AA) low  resolution ($\sim$ 6\,\AA) spectra were acquired during 
consecutive 16\,h with exposure times
equal or twice the $\rm P_{spin}$ to smear out effects of
the rotational pulsation (Table\,1). The SWP and LWP exposures sample 
the orbital period.

The spectra have been re-processed at VILSPA using the IUE NEWSIPS pipeline 
used for the IUE Final Archive which applies the SWET extraction method as well 
as the latest flux calibrations and close-out camera sensitivity corrections 
(Garhart et al. 1997). Line-by-line images have been inspected for spurious 
features which have been identified and removed.

\subsection{ The optical photometry}

\noindent Optical photometry was conducted in the period October 
18--23\, 1995 
at the SAAO  0.75\,m telescope  and UCT Photometer employing
a Hamamatsu
R93402 GaAs photomultiplier. BVRI (Cousins) photometry was 
carried out on the first three nights performing  symmetric modules 
with integration times of 30\,s or 20\,s respectively for the B and I,
or V and R filters. The typical time resolution for the sequence
of all four filters was $\sim$120\,s, with interruption every 
$\sim$ 10\,min for sky
measurements. The times and
durations of individual runs are reported in Table\,1. The orbital period
has not been fully sampled.

Additional fast 
photometry in white light has been carried out on October 21 and 23, 1995 
using the same photometer, but
employing a second channel photomultiplier for monitoring of a nearby
comparison star. Continuous 5\,s integrations were obtained, with occasional
interruptions (every 15--20\,min) for sky measurements (lasting $\sim$ 
20--30\,s).
During the two nights, the observations were carried 
out for 4.3\,h and  2.5\,h respectively. 

The photometric data have been reduced in a standard manner with sky 
subtraction, extinction correction and transformation to the standard
Cousins system using observations of E-region standards obtained on the 
same night.

\section{The UV spectrum of FO\,Aqr}

The grand average UV spectra of FO\,Aqr as observed with FOS and IUE are shown 
in Fig.\,1 (upper panel), where the above flux difference is apparent.

The UV luminosity in the 1150--3200\,\AA\  IUE range is  
$6\times 10^{32}$\,\es assuming 
a distance of 325\,pc (Paper\,1), a factor 1.3 larger than during
previous UV observations in 1990. The optical photometry also indicates 
a brightening of 0.17\,mag between the two epochs, indicating
long term luminosity variations (see also sect.\,8.5)

The UV spectrum of FO\,Aqr is typical of magnetic CVs 
(Chiappetti et al., 1989; de Martino 1995) with strong emissions of 
N\,V \,$\lambda 1240$, Si\,IV \,$\lambda 1397$, C\,IV\, $\lambda 1550$, 
He\,II\, $\lambda \lambda 1640, 2733$ and Mg\,II\, $\lambda 2800$. The weaker
Si\,IV\, with respect to N\,V emission,  classes the IP nature
(de Martino 1995). Weaker emissions from lower 
ionization states of different species such as C\,III\, $\lambda 1176$, 
the blend of Si\,III\, $\lambda 1298$ multiplet, Si\,II$\, \lambda 1304$ and 
geocoronal O\,I$\, \lambda 1305$, 
Si\,III\, $\lambda 1895$, 
Si\,II\, $\lambda 1808$,  N\,IV\, $\lambda 1718$, N\,III]\, 
$\lambda 1747-1754$, Al\,III\, $\lambda 1855$ and Al\,II\, $\lambda 1670$, 
as well as He\,II\, $\lambda 2307$, possibly blended with C\,III 
$\lambda 2297$ , and He\,II $\lambda  2386$, are identified in the 
higher quality FOS spectrum. Also weak 
oxygen lines of O\,IV\, $\lambda$1343 and O\,V\, $\lambda$1371 are detected.
Some of these lines are also observed in the HST/FOS spectra of AE\,Aqr
(Eracleous \& Horne 1994), DQ\,Her (Silber et al. 1996) and PQ\,Gem 
(Stavroyiannopoulos et al. 1997). The presence of high ionization species
together with extremely weak emissions (E.W. $<$ 1\, \AA\ ) of lower
ionization species are characteristic of a higher ionization efficiency in IPs 
with respect to Polars (de Martino 1998). The line ratios N\,V/Si\,IV and 
N\,V/He\,II, when compared with photoionization models developed by Mauche
et al. (1997) are close to the predicted values for an ionizing blackbody
spectrum at 30\,eV. 

In contrast to the IUE spectra, the FOS data allow us to finally detect
the intrinsic \La\ $\lambda$1216 line. This appears to 
be composed of a relatively deep (E.W.=5.4$\pm$0.1\, \AA \,) absorption 
and a weak emission (E.W.=1.4$\pm$0.1\,\AA\ ). The center wavelength of 
the absorption
feature is however red-shifted by $\sim$4\,\AA\ with respect to rest 
wavelength and other emission line positions,  while the weak emission is 
blue-shifted at 1206\,\AA\ in the grand average spectrum. 
Discussion on the nature  of this feature is left until sects.\,5 and 7, 
however the \La\ absorption provides 
an upper limit to the hydrogen column density along the line of sight to
FO\,Aqr. 

A pure damping Lorentzian profile (Bohlin 1975) convolved with 
a 7\,\AA\ FWHM Gaussian has then been fitted to the \La\ absorption line 
(Fig. 1, bottom panel). 
The resulting neutral hydrogen column density is $\Nh=(5.0\pm1.5)\times 10^{20} \,
\rm cm^{-2}$. The residual from the fit shows an emission line with maximum
flux at $\sim 1215$\,\AA\ , probably geocoronal or intrinsic, 
with an excess of flux in the blue wing possibly due to emission of
Si\,III\, $\lambda$1206. 

The derived value for \Nh \, is consistent with the total 
interstellar  column density in the direction of FO\,Aqr as derived from 
Dickey \& Lockman (1990) and with the upper limit estimated from
X-rays (Mukai et al. 1994). Assuming an average gas-to-dust ratio 
(Shull \& van Steenberg 1985), this upper limit corresponds to a 
reddening of $\rm E_{B-V} \approx 0.1$. Although FO\,Aqr was already known 
to be negligibly reddened from IUE observations (Chiappetti et al. 1989),
an upper limit $\rm E_{B-V}=0.013 \pm$ 0.005 is derived from the absence
of the 
2200\,\AA\ absorption in the FOS data. This indicates that, despite 
the coincidence, most 
of the neutral absorption is unrelated to the interstellar dust and hence 
it is likely located within the binary system.

\begin{figure}
\begin{center}
\mbox{\epsfxsize=9cm\epsfbox{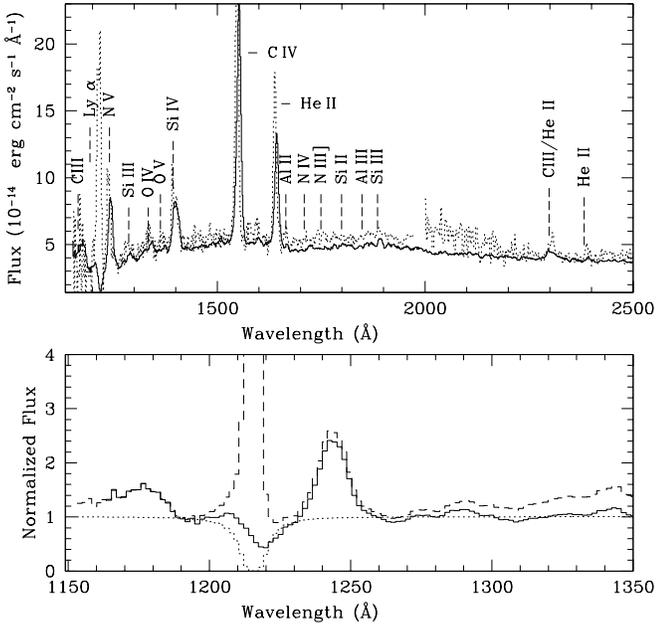}}
\caption[]{\label{average} {\bf Upper panel:} Grand average FOS (solid line) 
and IUE (dotted line) spectra of FO\,Aqr in the range 1150--2500\,\AA\ 
\, together with line identification.
{\bf Bottom panel:} Enlargement of the \La\ feature in the average FOS spectrum. 
The solid line represents the observed spectrum, the dotted line a damped 
Lorentzian profile convolved with a 7\,\AA\ FWHM gaussian corresponding to
$\Nh = 5 \times 10^{20}\,\rm cm^{-2}$ and the dashed line represents
the ratio between the two, enhancing the emission feature at 1215\,\AA\ .}
\end{center}
\end{figure}

\section{Time series analysis}

The presence of periodicities in FO\,Aqr has been investigated in the 
FOS continua, emission lines and zero-order light as well as in the 
optical photometric data. 

\subsection{HST UV data}

Fluxes in five line-free continuum bands have been measured in each FOS 
spectrum in the
ranges $\lambda$1265--1275, $\lambda$1425--1450, $\lambda$1675--1710, 
$\lambda$2020--2100, $\lambda$2410--2500. Line fluxes of 
He\,II $\lambda$1640, N\,V, Si\,IV and C\,IV and \La\
have been computed adopting a method which uses for the con\-ti\-nuum a
power law distribution as found from a fit in the above
continuum bands. Furthermore, since the low spectral resolution of FOS data
prevents the study of UV line profiles, measures of 
the V/R ratios of emission lines have been used to investigate possible 
motions in the lines. 
These are defined as the ratios between the integrated fluxes in the violet and
red portions of the emission lines assuming as centroid wavelength that
measured in the average profile.  Such analysis is restricted
to the strong emissions of He\,II, C\,IV and N\,V lines whose FWZI
are $\pm$3000\,\kms, $\pm$4000\,\kms and $\pm$3000\,\kms \,respectively.

\begin{figure*}[t]
\begin{center}
\mbox{\epsfxsize=15cm\epsfbox{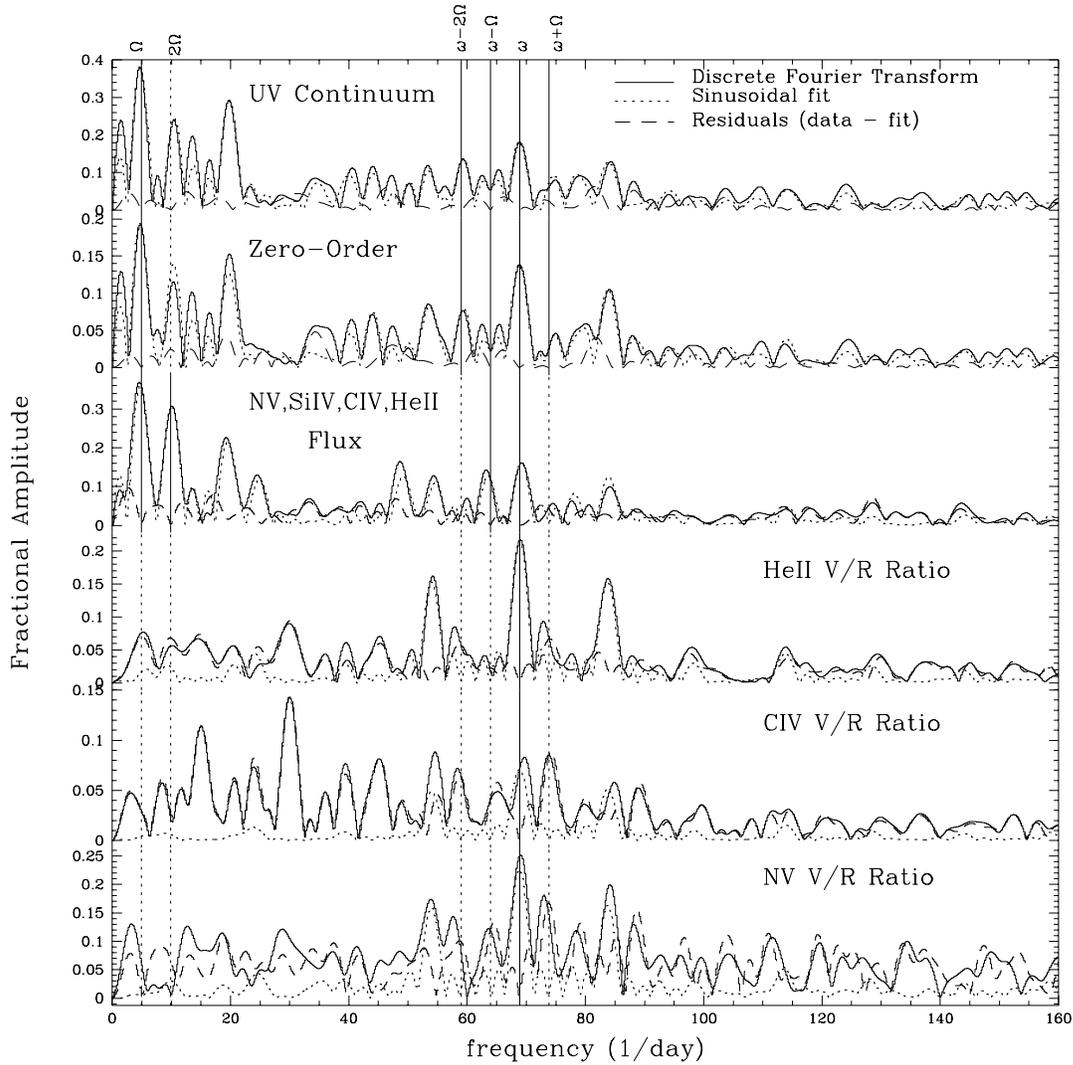}}
\caption[]{From top to bottom: DFTs of the UV continuum, zero-order 
light, emission line fluxes and V/R ratios of He\,II, C\,II and N\,V 
(solid line).
DFTs of the sinusoidal functions (dotted line) together with those of the
residuals (dashed line) are also shown.
The frequencies used in the multiple sinusoidal fits are marked with 
vertical solid lines.}
\end{center}
\end{figure*}

In order to detect the active frequencies in the power spectrum and to
optimize the S/N ratio,  a Fourier analysis has been performed using the 
DFT algorithm of Deeming (1975)  on the total UV continuum
(sum of the five bands) and line (sum of N\,V, Si\,IV, C\,IV and He\,II)
fluxes and zero-order
light. From Fig.\,2, the dominance of the $\Omega$ variability is apparent, 
being about twice the $\omega$ signal,  as well as the presence of 
substantial power  at the sideband and orbital harmonic frequencies. 
To distinguish real signals from artifacts due to the
sampling of the HST orbit, a least-square technique was applied to the
each data set which fits simultaneously
multiple sinusoids at fixed frequencies. A 
synthetic light curve with the same temporal sampling of the data
was created and subtracted (residuals). 
The continuum and zero-order light reveal, besides the $\omega$ and
$\Omega$ frequencies, also the
$\omega-2\,\Omega$, $\omega-\Omega$ and $\Omega+\omega$ sidebands, whereas
in the emission lines the $2\,\Omega$ and $\omega-\Omega$ frequencies are
detected. 
In Fig.\,2 the amplitude spectra relative to the multiple 
sinusoids and the residuals are also shown for comparison.
It should be noted that peaks at 
$\rm \sim 54\,day^{-1}$ and $\rm \sim  84\,day^{-1}$ are identified as 
sideband fequencies of the spin and HST orbital frequencies. These
are removed by the method as shown by the residuals.
Then a  five (four) frequency composite sinusoidal function 
for each spectral band  (emission 
line) has been used and  the derived amplitudes, reported in Table\,2,
 have been compared 
with the average power in the DFTs of the residuals ($\sigma$) in the 
range of frequencies of interest (i.e. $\nu \la$ 1.4\,mHz), (column 8).
While the signals at $\omega-2\,\Omega$ (continuum) 
and at $2\Omega$ (emission lines) on average fulfil a 4 $\sigma$ criterium,
the other sidebands are between between 2.1 and 3.5 
$\sigma$. 

A further check has been performed using the CLEAN algorithm 
(Roberts et al. 1987) which removes the windowing effects of the HST orbit.
The CLEANED power spectra, adopting a gain of 0.1 and 500 iterations, 
indeed reveal the presence of the 
$\omega -2\Omega$ and $\omega - \Omega$ sidebands in the continuum and the 
$2\Omega$ in the emission lines. The lack of significant power at 
the other frequencies is consistent with the previous analysis. 
Hence, these weakly active frequencies will be considered with some caution
in this analysis.

\begin{table*}[ht]
\begin{center}
\caption[]{Amplitudes of the modulations for the UV continuum, optical and
UV emission lines derived from a multicomponent sinusoidal fit.}
\begin{tabular}{lrrrrrrrr}
\hline
& & & & & & & & \\
\multicolumn{1}{c}{ Band (\AA)} &
\multicolumn{1}{c}{\bf $A_{\Omega}$} &
\multicolumn{1}{c}{\bf $A_{2\Omega}$} &
\multicolumn{1}{c}{\bf $A_{\omega}$} &
\multicolumn{1}{c}{\bf $A_{\omega-2\Omega}$} &
\multicolumn{1}{c}{\bf $A_{\omega-\Omega}$} &
\multicolumn{1}{c}{\bf $A_{\omega+\Omega}$} &
\multicolumn{1}{c}{\bf $\bar{A}_{res}$} &
\multicolumn{1}{c}{\bf $A_{\omega}(V/R)$} \\
& & & & & & & & \\
\hline
& & & & & & & & \\
{\bf 1265\,--\,1275} & 1.730(52) & 
                 & 0.893(51) & 0.913(72) & 0.391(55) & 0.693(72)
                 & 0.181 & \\
{\bf 1425\,--\,1450} & 1.784(42) &
                 & 0.867(43) & 0.694(59) & 0.453(44) & 0.441(61)
                 & 0.166 & \\
{\bf 1675\,--\,1710} & 1.779(37) &
                 & 0.738(37) & 0.685(51) & 0.344(37) & 0.420(52)
                 & 0.157 & \\
{\bf 2020\,--\,2100} & 1.416(29) &
                 & 0.656(29) & 0.553(40) & 0.303(30) & 0.334(41)
                 & 0.125 & \\
{\bf 2410\,--\,2500} & 1.130(25) &
                 & 0.627(25) & 0.463(34) & 0.236(25) & 0.257(34)
                 & 0.108 & \\
{\bf 2900\,--\,2985} & 0.942(45)&           &           & 
                 &       & \\

{\bf Zero-Order} & 0.578(15) &
                 & 0.417(14) & 0.323(21) & 0.120(15) & 0.178(21)
                 & 0.062 & \\

& & & & & & & & \\
\hline
& & & & & & & & \\

{\bf B}    & 0.354(25) &
                 & 0.259(38) & 0.025(33) & 0.046(35) & 0.047(31)
                 & 0.045 & \\
{\bf V}    & 0.182(14) &
                 & 0.129(22) & 0.050(19) & 0.006(21) & 0.023(18)
                 & 0.026 & \\
{\bf R$_{c}$} & 0.134(13) &
                    & 0.091(20) & 0.026(16) & 0.020(21) & 0.037(16)
                    & 0.021 & \\
{\bf I$_{c}$} & 0.109(10) &
                    & 0.054(16) & 0.022(14) & 0.031(17) & 0.015(13)
                    & 0.016 & \\
& & & & & & & & \\
\hline
& & & & & & & & \\
{\bf N\,V flux} & 14.37(1.98) & 6.75(1.97) & 12.75(1.37)
                & & 12.25(1.37) &
                & 3.32
                & 0.256(41) \\
{\bf Si\,IV flux} & 15.40(1.54) & 13.17(1.53) & 7.60(1.07)
                  & & 7.77(1.07) &
                  & 2.75
                  & \\
{\bf C\,IV flux} & 77.10(2.87) & 36.74(2.86) & 35.71(1.98)
                 & & 22.22(1.99) &
                 & 16.99
                 & 0.057(08) \\
{\bf He\,II flux} & 31.81(1.59) & 14.68(1.60) & 22.57(1.10) 
                  & & 13.40(1.11) &
                  & 2.75
                  & 0.173(10) \\
{\bf Ly$\alpha$ abs.}       & 23.35(1.08) & & 6.83(1.07) & & & & 3.41  & \\    
& & & & & & & & \\
\hline
\end{tabular}
\end{center}
Notes: (1) Continuum flux amplitudes are in units of $10^{-14}$\,\ecsa 
while line fluxes are in units of $10^{-14}$\,\ecs. 
Errors in parentheses are referred to the last significant digits.\\
(2) $\bar{A}_{res}$ is the average amplitude of the DFT of the residuals,
calculated for frequencies $\la$1.4\,mHz.\\
(3) For the $\lambda \lambda$ 2900--2985 band only the orbital amplitude is
reported as derived from IUE data.\\
\end{table*}

A strong colour effect in the UV continuum  is detected with amplitudes 
decreasing at longer wavelengths.
Different from other lines is the behaviour in the \La\ feature, whose 
absorption component 
is modulated at the $\Omega$ frequency, while a variability at the 
spin is at a  2$\sigma$ level. No  significant variations are detected in the
equivalent width of the absorption as well as in the emission component. 

In contrast to the flux behaviour,  the V/R ratios are variable only at the $\omega$ frequency 
 (Fig.\,2, bottom panels).  Noteworthy is the marginal spin variability in the 
C\,IV line. The amplitudes of the spin modulation, obtained 
from the least-square fits, are reported in the last column of Table\,2.

\subsection{Optical data}

The analysis of BVRI data acquired during three nights 
has been performed following the
same procedure adopted for the HST data. Contrary to 
previous observations, the orbital variability also dominates in the optical
being  $\sim$ 1.5 times the spin modulation.
The presence of sideband modulations is more uncertain because of the 
lower quality of the data. 
Nevertheless, to ensure uniformity between UV and optical results,
a least-square fit to the data has been applied using the same five 
frequency sinusoidal function, i.e. $\Omega$, $\omega$, $\omega-2\,\Omega$,
$\omega-\Omega$, $\omega+\Omega$.
The resulting amplitudes, reported in Table\, 2, when compared with the
noise in the residuals (column 8), can be considered as upper limits.

As far as the fast photometry is concerned, the low quality of the
data only allows the detection of the spin and orbital variabilities.
In particular, the latter is detected on the first night with a pronounced
dip which is not consistent with the refined
orbital ephemeris based on orbital minima recently given by Patterson et al. 
(1998), 
which defines the inferior conjunction of the secondary star. 
Therefore these data will not be used for a multi-wavelength analysis of 
the pulsations.

>From this analysis new times of maxima for the orbital and 
rotational modulations are derived for the UV continuum and optical light:

\begin{description}
\item
$ HJD^{\rm max}_{\rm orb} = 2\,449\,971.2251 \pm 0.0006$ in the UV;\\
\item 
$ HJD^{\rm max}_{\rm spin} = 2\,449\,971.11181 \pm 0.00010 $ in the UV;\\
\item
$ HJD^{\rm max}_{\rm orb} = 2\,449\,971.2235 \pm 0.0034$ in the optical;\\
\item
$ HJD^{\rm max}_{\rm spin} = 2\,449\,971.11021 \pm 0.00035 $ in the optical;\\
\end{description}

Both UV and optical orbital maxima lead by $\Delta \Phi_{\rm orb}$ = 0.145 
those predicted by Patterson et al.'s ephemeris. Such phase difference, 
discussed in  sect. 7, is consistent with the previous UV results (Paper\,1). 

On the other hand, the 
optical rotational maximum agrees within 8 per cent with that
predicted by the new revised cubic ephemeris given by Patterson (1998, private
communication):\\

\begin{center}

$ HJD ~~ \, 2\,444\,782.9168(2) + 0.014519035(2)\,E +$ 
\end{center} 

\begin{flushright} $ 7.002(7)\,10^{-13}\,E^{2} -  1.556(2)\,10^{-18}\, 
E^{3}$ ~~~~~~~~~~~~~~~~~~~~~~ (1) \end{flushright}

The UV rotational pulses lag the optical by 
$\Delta \Phi_{\rm rot} =0.186$. Such colour effect will be discussed in 
more detail in sect.\,5.

Furthermore, the time of coherence between spin and beat modulations is 
found to be $HJD_{\rm co}$ = 2\,449\,971.2335 $\pm$ 0.0070 in both 
UV and optical. Throughout this paper the Patterson et al.'s orbital 
ephemeris will be used but 
$\Phi_{\rm orb}$ = 0.0 will refer to the orbital maximum. Hence, phase 
coherence occurs at $\Phi_{\rm orb} = 0.90\pm$0.03. i.e.
close to the orbital maximum, whilst in 1988 and 1990 
it was found close to the orbital minimum (Osborne \& Mukai 1989; 
Paper\,1). Such phase changes are not uncommon for FO\,Aqr (Semeniuk \& 
Kaluzny 1988; Hellier et al. 1990).

\section{The rotational modulation}

UV continuum, zero-order light, and UV emission line fluxes 
as well as their V/R ratios have been folded in 56 phase bins along 
$\rm P_{spin}$. The spin pulses, 
pre-whitened from all other 
frequency variations, are shown in Fig.\,3 together with the optical B band 
pulses folded in 28 phase bins.  A strong colour effect is observed
in both amplitudes and phasing. Fractional amplitudes (amplitudes of 
the sinusoid $\rm A sin (\omega t + \phi)$ divided by the average value) 
decrease from
26$\%$ in the far-UV to $16\%$ in the near-UV and to $10\%$ in the optical. 
The far-UV pulse maximum is broader and lags by $\Delta \phi_{\rm spin} = 0.091\pm
0.010$ and 0.186$\pm$0.008  the near-UV and
optical maxima respectively. 
While the UV line fluxes follow the near-UV modulation,
their V/R ratios are in phase with the far-UV, with a maximum blue shift
when the far-UV pulse is at maximum. The pulsation in the line
fluxes are 18$\%$, 12$\%$, and 10$\%$ in He\,II, C\,IV and Si\,IV. The
V/R ratios are generally $<$1 indicating the presence of a dominant 
blue component in the lines, which is also visible from the extended blue wings
in the average spectrum.
The velocity displacements indicate the
presence of a spin S-wave in the line profiles similar to the optical 
(Hellier et al. 1990). Both continuum and emission lines therefore strongly 
indicate the presence of two components which affect the rotational 
modulation in FO\,Aqr.

\begin{figure*}[t]
\begin{center}
\mbox{\epsfxsize=16.0cm\epsfbox{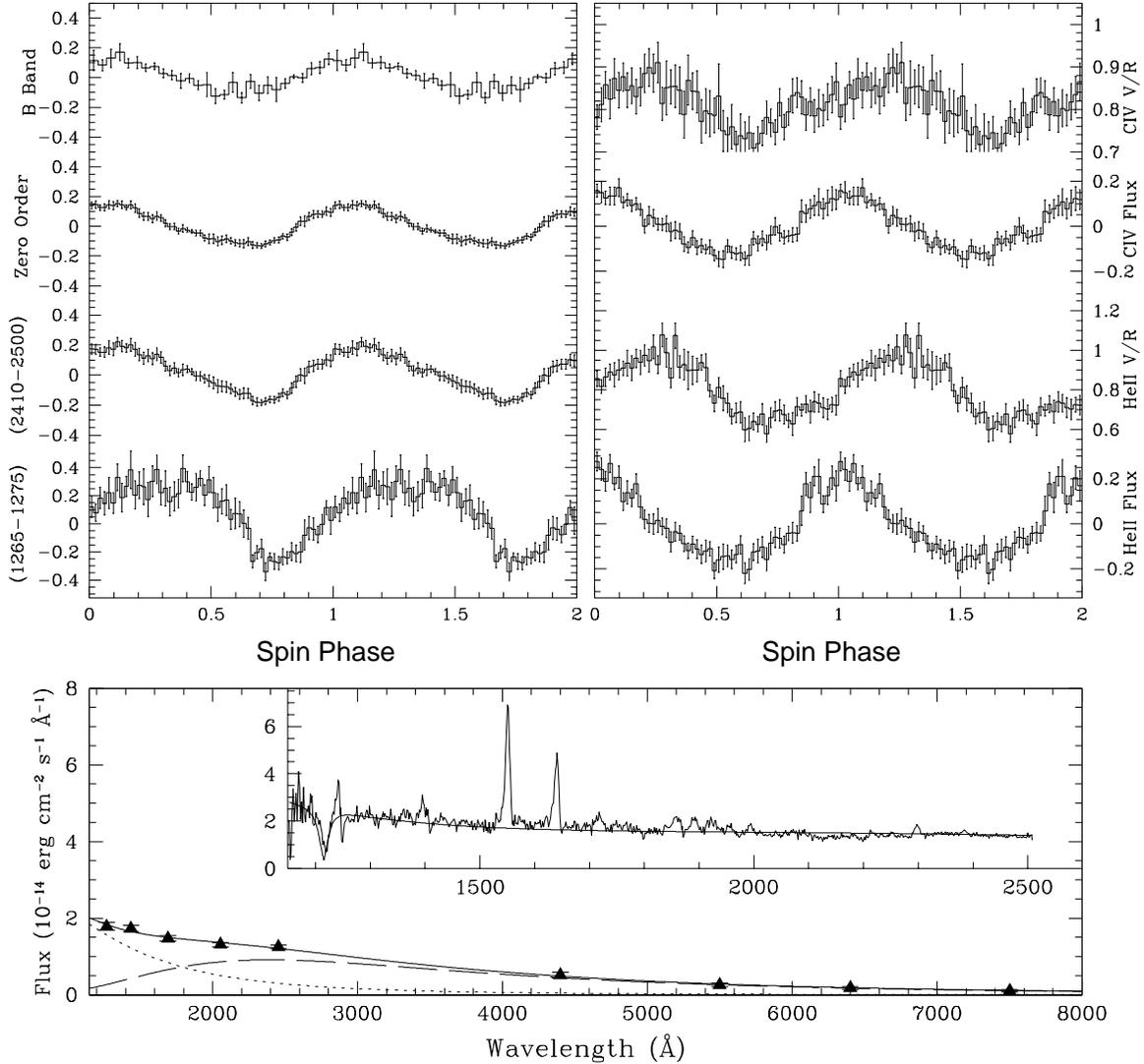}} 
\caption{{\bf Upper left:} Spin pulses in the far-UV, near-UV, zero order 
and B band
(upper left panel). {\bf Upper right:} C\,IV and He\,II flux and V/R 
rotational curves. Fluxes are fractional as described in the text,
the average value has been subtracted. {\bf Bottom panel:} Rotational
broad band UV and optical energy distribution together with the best fit 
composite function: a hot 
(37\,500\,K) (dotted line) and a cool (12\,000\,K) (long-dashed  line) blackbody 
function (sum: solid line). A composite function (solid line) 
consisting of a 36\,000\,K white 
dwarf model spectrum and of the same 12\,000\,K blackbody absorbed by 
$\Nh = 5 \times 10^{20}\, \rm cm^{-2}$ 
is shown in the insert figure together with the FOS rotational pulsed spectrum
described in the text.}
\end{center}
\end{figure*}

\noindent The 797 FOS spectra have been spin-folded into 20 phase bins. A total 
of  780 light curves, each sampling a wavelength 
bin of $\sim$ 1.8\,\AA\ , were then produced and fitted  with a sinusoid.
The resulting amplitudes define
the rotational pulsed spectrum as $\rm F_{\lambda} = 2\,A_{\lambda}$.
This spectrum (shown in the enlargement of the lower panel of Fig.\,3), 
has to  be regarded as an upper limit to the modulated flux since 
no pre--whitening could be performed given the low S/N of each 780 light 
curves. This spectrum gives evidence of modulation not only in the
main emission lines and \La\ absorption but also in the weaker emission 
features identified in sect.\,3. Broad band UV continuum
and optical photometric spin pulsed fluxes, obtained from the 
multi-frequency fit and reported in Fig.\,3, provide a correct 
description of the rotational pulsed energy distribution. 
A spectral fit to the broad band UV and optical spectrum, using 
a composite
spectral function, consisting of two blackbodies, gives
37\,500$\pm$500\,K and 12\,000$\pm$400\,K
($\chi^{2}_{\rm red} = 0.91$). The 
projected fractional area of the hot component is 
$\sim$0.11\,A$_{\rm wd}$, while the cool one covers
$\rm \sim 10.7\,A_{\rm wd}$, for $\rm R_{wd}=8\times 10^{8}\,cm$, 
and d=325\,pc (Paper\,1). 

The FOS rotational pulsed spectrum shows the pre\-sen\-ce of a \La\ absorption 
feature which gives  a hydrogen column density of 
$\rm 8 \pm 2 \times 10^{20}\,cm^{-2}$. On the other hand, 
assuming $\Nh = 5 \times 10^{20}\, \rm cm^{-2}$, as derived
from the grand average spectrum, a composite function consisting of a white
dwarf model atmosphere with at 36\,000\,K and of the same 12\,000\,K blackbody  
gives an equally satisfactory fit to the whole FOS spectrum.
For a distance of 325\,pc the radius of the white dwarf is
4.9$\times 10^{8}$\,cm, in agreement with that of the hot blackbody component.

\noindent While the detection of the hot component is new, a comparison with 
previous spectral analysis of the optical and IR
spin pulses observed in 1990 shows that the temperature of the
cool component has not changed with time but instead it suffered a decrease
in area by a factor of $\sim$ 1.5.

\section{The sideband modulations}

 The beat $\omega - \Omega$ modulation, although weak among the 
detected sidebands, shows an anti-phased behaviour between
line fluxes and continuum (Fig.\,4a). The UV line fluxes, 
with average fractional amplitudes of $\sim 11\%$, 
show a minimum when the UV continuum is maximum. Their maximum shows
a dip-like feature centered on the UV continumm minimum.
Colour effects are also encountered, the far-UV pulses being stronger
($\sim 11\%$) than the near-UV ($\sim 6\%$) ones and lagging by $\sim$ 0.2 in
phase the near-UV. The trend of a decrease in the amplitudes at increasing 
wavelengths  is confirmed in the optical where an 
upper limit of $\sim$ 2$\%$ can be set to the fractional amplitudes.

\begin{figure*}[h,t]
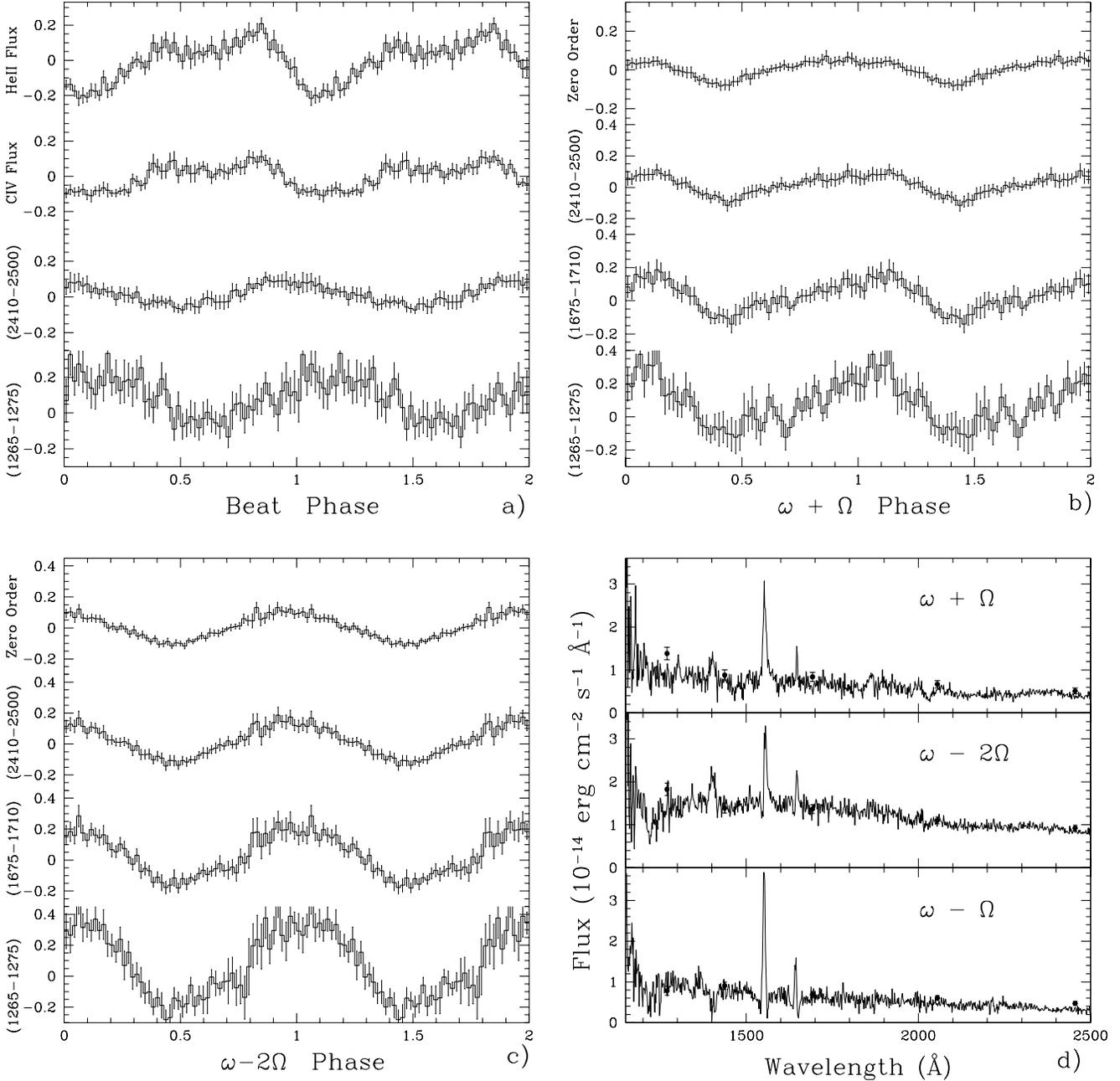

\begin{center}
\mbox{\epsfxsize=8.8cm\epsfbox{8955.f4a}} 
\mbox{\epsfxsize=8.8cm\epsfbox{8955.f4b}} 
\mbox{\epsfxsize=8.8cm\epsfbox{8955.f4c}} 
\mbox{\epsfxsize=8.8cm\epsfbox{8955.f4d}} 
\caption{{\bf a} The beat $\omega - \Omega$ modulation in the far-UV, near-UV
continua and line fluxes of C\,IV and He\,II.  {\bf b} The $\omega
+ \Omega$ and {\bf c } the  $\omega - 2\,\Omega$ light curves in three 
UV bands and zero order light. Fluxes are fractional. {\bf d} From bottom
to top, the 
pulsation spectra  at the beat $\omega - \Omega$, $\omega - 2\,\Omega$ 
and $\omega + \Omega$ frequencies (solid line) together with broad band 
UV fluxes derived from the multi-frequency fits (filled circles).}
\end{center}
\end{figure*}

\noindent A similar colour effect is also observed in the 
$\omega + \Omega$ sideband pulsation with fractional amplitudes ranging from 
$\sim 20\%$ in the far-UV to $\sim 7\%$ in the near-UV, displaying 
a broadening of the maximum towards longer wavelengths (Fig.\,4b). 

\noindent The strong $\omega - 2\,\Omega$ pulsation only
shows a wavelength dependence in the fractional amplitudes, being 
in the far-UV $\sim 26 \%$ and decreasing to $\sim 12\%$ in 
the near-UV (Fig.\,4c) and to $\sim 2-3\%$ (upper limit) 
in the optical.


\noindent The modulated spectra at the three frequencies have been
derived following the same procedure as for the rotational pulsed spectrum.
The anti-phased behaviour of the emission lines is seen in the beat
pulsed spectrum, where these are seen as absorption features, except for 
residuals
in the C\,IV and He\,II lines (Fig\,4d, bottom panel). The modulation 
spectra at 
$\omega - 2\Omega$ and $\omega +\Omega$ frequencies show weak emissions
at Si\,IV, C\,IV and He\,II indicating a marginal variability at these
frequencies.  The UV continuum  energy distributions of these variabilities 
are best represented 
by power laws $\rm F_{\lambda} \propto \lambda ^{-\alpha}$ with spectral index 
$\alpha \sim 0.8-1.7$, rather than blackbodies (20\,000 -- 25\,000\,K), possibly
suggesting that more than one component is acting. Given the low level of
confidence of the sideband variabilities it is not possible to derive 
further information.

\section{The orbital variability}

The UV and optical orbital modulations have been investigated
folding the FOS continuum broad band and emission line fluxes in 28
orbital phase bins. 
The light curves have been prewhitened by the other active frequencies using
the results of the multi-frequency fits. For the IUE 
data, continuum broad
band and emission line flux measures have been performed on each SWP and 
LWP spectrum. Three broad bands have
been selected in each spectral range, five of them coinciding with the FOS
selected bands and a sixth one in the range $\lambda$ 2900--2985. 
The contribution of the spin pulsation, as derived from the multi-frequency
fit has been removed. The best fit blackbody spin pulsed spectrum 
has been used to allow prewhitening in the range $\lambda$2900--2985.
The FOS and IUE broad band continuum fluxes in the far-UV, mid-UV and near-UV 
as well as the zero order and B band light curves are reported in the left 
panel of Fig.\,5,  while the emission line fluxes of Si\,IV, C\,IV and He\,II
are shown in the right panel. The orbital gaps due to the HST sampling
are apparent.

\begin{figure*}[h,t]
\begin{center}
\mbox{\epsfxsize=16.0cm\epsfbox{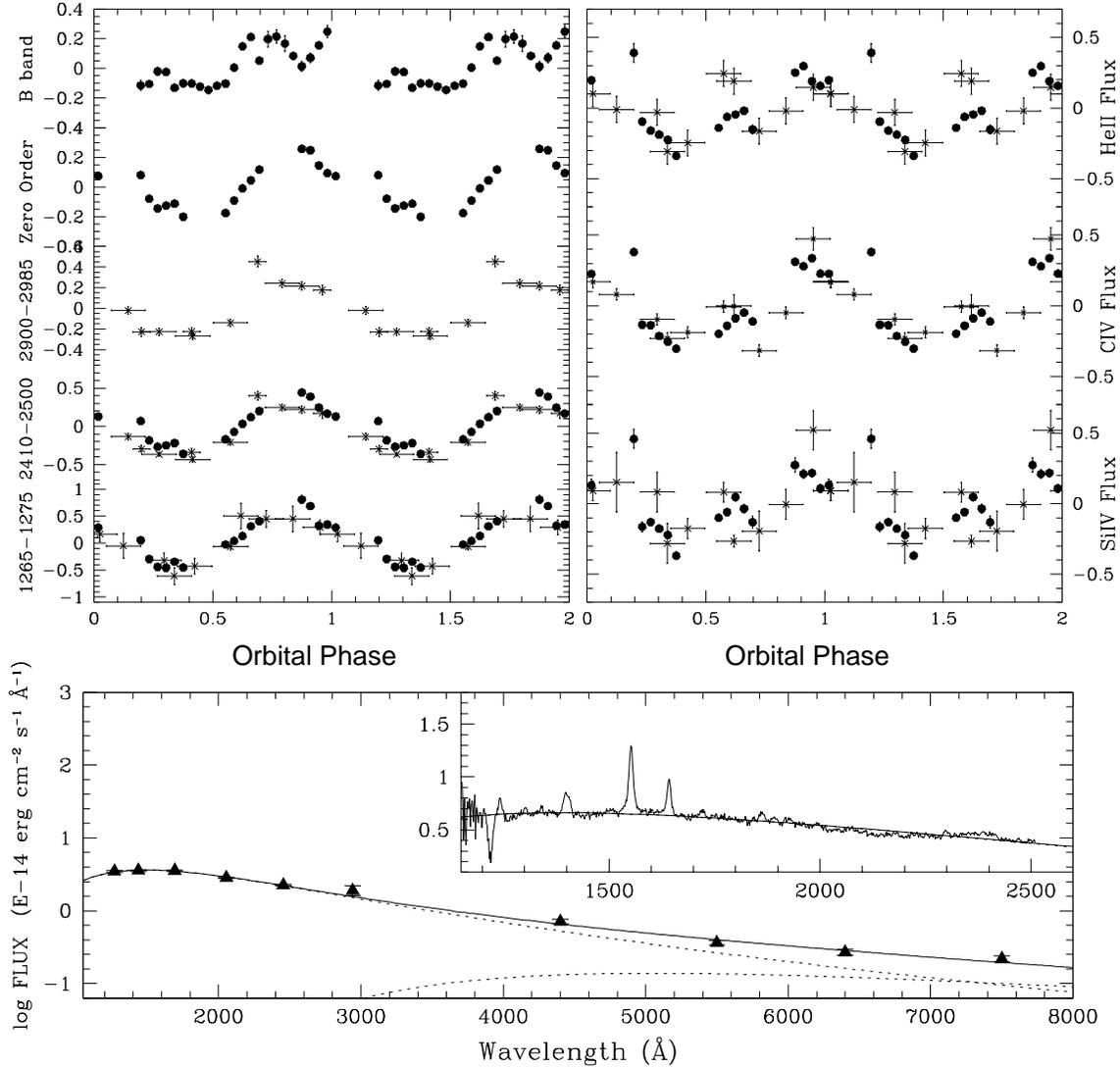}} 
\caption{{\bf Upper left:} Orbital modulation in the far-UV, mid-UV and near-UV,
 zero order and B light
(upper left panel). {\bf Upper right:} Si\,IV, C\,IV and He\,II flux light
curves. Fluxes are fractional as described in the text
 and the average value has been subtracted. The IUE measures are reported 
together with their phase coverage. {\bf Bottom panel:} The orbital
broad band UV and optical 
modulated fluxes represented with the best fit (solid line) composite 
function as described in the text. The two blackbodies are represented with 
dotted lines. A hot (21500\,K) blackbody function describes the UV FOS
spectrum (shown in the inserted figure.)}
\end{center}
\end{figure*}

A strong colour dependence is encountered in the  modulation amplitudes as 
well as the phasing.   Fractional amplitudes 
range from 40$\%$ in the far-UV to 28$\%$ in the near-UV (IUE band)
and 15$\%$ in the optical. The modulation amplitudes then have increased
by a factor $\sim$ 2.5 in the UV and $\sim$ 1.5 in the optical with respect 
to 1990. However, the phasing of UV maximum and minimum has 
not changed with time, occurring at
$\rm \Phi_{orb}=0.86$ and $\rm \Phi_{orb}=0.34$, respectively.
The UV modulation is more sinusoidal, whilst the optical
light curve is more structured with a double humped maximum between 
$\rm \Phi_{orb}$ = 0.75 and $\rm \Phi_{orb}$ = 0.0.  
A comparison with the optical behaviour in 1990 indicates an absence
of a broad maximum centered at $\rm \Phi_{orb}$ = 0.0 
and a less defined minimum. The current observations
are inadequate to resolve the orbital dip due to a grazing eclipse of
the accretion disc in either UV and optical ranges. 

\noindent The orbital modulation in the UV 
emission line fluxes is strong with fractional amplitudes of 
40$\%$ in N\,V, 20$\%$ in Si\,IV, 29$\%$ in C\,IV and 23$\%$
in He\,II and almost in phase with the UV continuum.

The spectrum of the UV orbital variability derived with the same procedure
as described before is shown in the enlargement of Fig.\,5 (bottom panel). 
>From the inserted figure
the \La\ absorption feature is apparent and is consistent 
with the neutral hydrogen column density of 
$8\pm 2 \times 10^{20}$\,cm$^{-2}$ inferred from the spin pulsed spectrum.
Hence, while this absorption in the orbital pulsation spectrum is clearly 
circumstellar, the same nature in the spin pulsed spectrum cannot be 
excluded.

The UV FOS spectrum 
requires a hot component with a blackbody temperature of
21\,500$\pm$500\,K (Fig.\,5 bottom panel enlargement).
The composite UV and optical broad band energy distrubution confirms the
previous results on the presence of two components, a hot 
at 19\,500 $\pm$ 500\,K and a cool one at 5\,700 $\pm$ 200\,K 
($\chi^{2}_{\rm red} = 1.1$) (Fig.\,5, bottom panel). With current UV observations, it is 
now possible to constrain
the temperature of the hot emitting region.  The temperature of 
the cool component is in agreement, within errors, with that
inferred in Paper\,1. A substantial increase by a factor of 
$\sim 2.6$ in the area of the hot region is found when compared to the 1990
epoch, which is  $\rm \sim 12\,A_{wd}$. The emitting area of 
the cool component is instead similar to that previously derived (Paper\,1).

\section{Discussion}

The HST/FOS and IUE spectroscopy has revealed new insights in 
the UV variability of FO\,Aqr. 

\subsection{The periodic variations}

The UV continuum and emission line fluxes are found to be strongly
variable at the orbital period. This
periodicity also dominates the optical range where FO\,Aqr was previously
found to be spin dominated. The time series analysis 
indicates the presence of other periodicities, 
the negative $\omega - 2\,\Omega$ sideband being much stronger than the beat
$\omega - \Omega$ in both UV and optical ranges. 
The presence of sidebands with different amplitudes at different
epochs is rather common in FO\,Aqr, although the beat is usually the strongest
(Patterson \& Steiner 1983; Warner 1986; 
Semeniuk \& Kaluzny 1988,  Chiappetti et al. 1989; Paper\,1, 
Marsh \& Duck 1996). In particular the intermittent occurrence of a stronger
pulsation at $\omega - 2\,\Omega$ frequency  was already noticed (Warner 1986;
Patterson et al. 1998). The strong negative sideband 
$\omega -2\,\Omega$, cannot be 
produced by an  amplitude modulation at $2\Omega$ frequency of the 
rotational pulses, since the positive sideband $\omega + 2\,\Omega$ should have
been present. Also, an orbital variability of the amplitude of 
the $\omega - \Omega$ modulation cannot be responsible alone since it is
too weak. Hence the $\omega - 2\Omega$ pulsation should be dominated by the 
effects of an unmodulated illumination from the white dwarf, which naturally
gives rise to the orbital variability (Warner 1986). 

\noindent The occurrence of coherence between spin and beat pulsations 
appears to be different from epoch to 
epoch (Semeniuk \& Kaluzny 1988; Osborne \& Mukai 1989; Paper\,1). 
It was proposed that phase coherence close to the optical orbital
minimum could be possible if the reprocessing site(s) are viewing the lower 
accreting pole (Paper\,1). The observed shift of half an orbital cycle would 
then imply that the reprocessing  region(s), are now viewing the main 
accreting pole, as predicted by the standard reprocessing scenario 
(Warner 1986).

\noindent The behaviour of UV emission lines is different between 
fluxes 
and V/R ratios. The line fluxes are strongly
variable at the orbital period, the spin variability being 1.6\, times
lower. On the other hand, their V/R ratios only show a rotational S-wave, 
but that of C\,IV line is surprisingly weak. The lack of detection of an 
orbital S-wave, can be ascribed to the low amplitude ($\sim$ 300--400\,\kms)
velocity displacements known from optical data (Hellier et al. 1989; 
Marsh \& Duck 1996), which are not detected because of the low spectral resolution
of the FOS data.

\subsection{The rotational pulses}

Both shapes and amplitudes of UV and optical continnum spin pulses indicate
the presence of two components, one dominating the near-UV and optical
ranges, already identified in Paper\,1 and a new contribution
dominating the far-UV pulses which lags by $\sim$ 0.2 in phase the first one.
Furthermore a different behaviour between emission line fluxes and 
V/R ratios is observed. While the latter show a spin S-wave in 
phase with the far-UV continuum, 
the line fluxes follow the near-UV and optical pulsations. 
The maximum blue-shift found at rotational maximum of the far-UV pulses 
indicates that the bulk of velocity motions in the emission lines 
maps the innermost regions of the accretion curtain. The outer curtain 
regions are then responsible for X-ray illumination 
effects seen in the line fluxes and near-UV and optical continua.
A direct comparison with 
previous X-ray observations reported by Beardmore et al. (1998) is not 
possible since, adopting their linear spin ephemeris, the UV and optical 
maxima lag by $\Delta \Phi_{\rm spin}$=0.4 
their predicted optical  maximum. However, the X-ray pulse maxima observed
by Beardmore et al. (1998), typically lag by $\Delta \Phi_{\rm spin}$ =0.2 
their optical phase zero (see their Fig.\,3), 
consistently with the lag observed between the far-UV and optical pulses. Hence
this difference is an indication that the far
sides of the accretion curtain come into view earlier than the innermost regions.

The spectrum of the pulsation reveals 
regions at $\sim$ 37\,000\,K 
covering a relatively large area, $\sim 0.1\,\rm A_{wd}$, with respect to 
typical X-ray fractional areas $f < 10^{-3}$ (Rosen 1992). 
Such hot components have been also observed
in the IPs PQ\,Gem (Stavroyiannopoulos et al. 1997) and EX\,Hya
(de Martino 1998). The presence of the  \La\ absorption feature 
in this spectrum can be partially due to the photospheric absorption of the 
heated white dwarf with similar temperature and fractional area as a
blackbody representation. On the other hand, both orbital and rotational
modulated spectra give similar values of $ \Nh $ if this absorption is of 
circumstellar nature. Hence, 
while only in AE\,Aqr the UV pulses are clearly associated
with the heated white dwarf (Eracleous \& Horne 1994, 1996), those in
FO\,Aqr can be associated with either the innermost regions of the accretion
curtain onto the white dwarf or its heated polar regions.

The second component 
identified as a cool 12\,000\,K region covers $\sim 11\rm A_{wd}$,
a factor $\sim 1.5$ lower than previously found in Paper\,1. 
Thus the decrease in the optical amplitudes does not involve substantial
changes in 
temperatures but in the size of 
the accretion curtain.
Such lower temperatures characterizing the near-UV and optical pulses are 
also recognized  in other IPs (de Martino et al. 1995; Welsh \& Martell 1996; 
Stavroyiannopoulos et al. 1997; de Martino 1998). Although a
two component pulsed emission might be a crude representation, it is
clear that temperature gradients are present within the accretion curtain
extending up to $\sim 6 \rm R_{wd}$.

The bolometric flux involved in the spin modulation due to both 
components amounts to 6.3$\times 10^{-11}$\,\ecs. Although no
contemporary X-ray observations are available this accounts for 
$\sim 26\%$ the total accretion luminosity as derived from ASCA 1993 
observations (Mukai et al. 1994).

\subsection{The sidebands variability}

The UV continuum pulsations observed at the sideband frequencies, 
$\omega - 2\Omega$,
$\omega - \Omega$ and $\omega + \Omega$, indicate the presence
of a relatively hot component $\sim$ 20\,000-25\,000\,K. The lack of adequate
data in the optical range does not allow one to confirm the cool 
($\sim$ 7\,000\,K), and hence possible second component, in the beat
pulsed energy distribution as found in Paper\,1.    The phase lags of 
the far-UV maximum with respect to that in the near-UV in 
the $\omega - \Omega$ and $\omega + \Omega$ pulsations are similar to
that observed in the rotational pulses. This is  consistent with the 
pulsations being produced by amplitude variations 
of the spin pulses at the orbital period. In contrast, the prominent 
negative sideband $\omega - 2\Omega$ variability is not 
affected by phase shift effects, indicating that indeed such
variability is mainly due to an aspect dependence of the reprocessing
site at the orbital period.

\noindent No strong pulsation in the UV emission lines 
is observed at these frequencies
except for the interesting anti-phased behaviour of these lines at the
beat period. These are observed as weak absorption features in the
modulated spectrum. Such behaviour, although much more prominent,  is also 
observed in the IP PQ\,Gem (Stavroyiannopoulos et al. 1997). Though
this is not easy to understand, a possibility could be that the 
reprocessing site, producing the $\omega - \Omega$ component in the
emission lines, is viewing the lower pole instead of the main X-ray
illuminating pole as also suggested by Stavroyiannopoulos et al. (1997).

\subsection{The orbital variability}

The present study confirms previous results where
the orbital modulation is composed by two contributions, identified
as the illuminated bulge and the heated face of the secondary star, the former
being at superior conjunction at $\Phi_{\rm orb}$ = 0.86, while the
latter is at superior conjunction at $\Phi_{\rm orb}$ = 0.0. The 
double-humped maximum in the optical light curve can be understood
in terms of relative proportion of a strong bulge contribution
with respect to that of the secondary star. 
Indeed  no changes 
in the temperatures are found (the hot one is better constrained with the
present data), but a substantial change by a factor 
of $\sim 2.6$ since 1990 is found in the emitting area of the bulge itself. 
All this indicates that 
illumination effects are basically unchanged whilst 
the inflated part of the disc has increased. 

The total bolometric flux involved in the orbital variability amounts
to 2.7$\times 10^{-10}$\,\ecs which is a factor $\sim$ 4 larger than that of the
rotational pulsation. Neglecting the sideband contributions
at first approximation, the total modulated flux amounts to 
3.3$\times 10^{-10}$\,\ecs and corresponds to a reprocessed 
luminosity of $\rm 4.2 \times 10^{33}$\,\es  which is approximately the order
of magnitude of the accretion luminosity derived from X-rays (Mukai et al. 1994).
Then assuming the balance of the energy budgets of the reprocessed 
and primary X-ray radiations, and hence of the accretion luminosity, 
an estimate of the accretion rate of 
$\rm \dot M = 6.7 \times 10^{-10}\, M_{\odot}\, yr^{-1}$ 
is derived.

\subsection{The long term variability}

FO\,Aqr has displayed a change in its power spectrum at optical and 
UV wavelengths on a time scale of five years. It was also brighter 
by 0.3\,mag and 0.2\,mag in the two ranges with respect to 1990.
This difference is accounted for by the orbital modulated flux.
The study of the spectra of the periodic variabilities has shown
that a shrinking of the accretion curtain by a factor of $\sim 1.5$ 
has occurred while the inflated part of the disc has increased in 
area by a factor of 2.6. Such changes indicate variations in 
the accretion parameters.  Worth noticing is that the unmodulated
UV and optical continuum component has not changed with time, indicating
that the steady emission from the accretion disc (Paper\,1) has not been 
affected. These results are in agreement
with the long term trend of the X-ray power spectra (Beardmore et al. 1998)
which showed that FO\,Aqr was dominated by the spin pulsation in 1990, while
in 1988 and in 1993 prominent orbital and sideband variabilities were present.
Changes from a predominant disc-fed accretion to
a disc-overflow (or stream-fed) accretion have been the natural explanation
for such changes in the X-ray power spectra. As proposed in Paper\,1,
the bulge provides a source for a variable mass transfer onto the
white dwarf, and an increase in its dimensions 
accounts for a predominant disc-overflow towards the
white dwarf.

Beardmore et al. (1998) suggested that changes in the accretion mode
could be triggered
by variations in the mass accretion rate and the analysis presented here
is indeed in favour of this hypothesis. An estimate of changes in the accretion 
rate producing a shrinking of the accretion curtain can be inferred
by the relation betwen magnetospheric radius, accretion rate and magnetic 
moment (Norton \& Watson 1989):

\begin{flushright}$ r_{mag} = \phi\, 2.7\times 10^{10}\, \mu_{33}^{4/7}\, \dot M_{16}^{-2/7}\, 
M_{wd}^{-1/7}$ ~~~~~~~~~~~ ~~~ ~~~~~~~~~~~~ (2) \end{flushright} 

where $\phi \la 1$ is a dimensionless factor accounting for a departure from
spherical symmetry,  $\mu_{33}$ is the magnetic moment in units of 
$\rm 10^{33}\,G\,cm^{3}$,
$\rm \dot M_{16}$ is the mass accretion rate in units of $\rm 10^{16}\,g\,s^{-1}$
and $\rm M_{wd}$ is the white dwarf mass in units of $\rm M_{\odot}$ . 
Hence assuming
that the accretion curtain reaches the magnetospheric boundary, a reduction 
of the linear extension by a factor of  $\sim$ 1.2 implies that the 
accretion rate has increased by a factor $\sim$ 2 in five years.

\noindent The long term spin period variations  observed from 1981 to 1997
in FO\,Aqr (Patterson et al. 1998), which changed from a spin-down to a 
recent spin-up since 1992, were proposed to be due to variations 
around the equilibrium
period produced by long term variations in 
$\rm \dot M$. The increase in brightness  level and the results found 
in the present analysis are strongly in favour  of this  interpretation.

\section{Conclusions}

The present HST/FOS and IUE spectroscopy has allowed us for the first
time to infer the characteristics of multiple periodicities in the 
UV emission of FO\,Aqr. Also, coordinated optical photometry has provided
further constraints summarized as follows: 

\medskip

{\it 1)}~~ The rotational pulsations in the UV and optical are 
consistent with the current accretion curtain scenario. They reveal the
presence of strong temperature gradients within the curtain moving 
from the footprints onto the white dwarf surface to several white dwarf radii.
A spin S-wave is observed in the UV emission lines which maps the innermost
regions of the accretion curtain.\\

{\it 2)}~~Orbital sideband variabilities indicate that reprocessing is occurring
at fixed regions within the binary frame. Different behaviour is
observed in the emission lines and continuum beat pulses indicating
a complex reprocessing  scenario.\\

{\it 3)}~~The orbital UV variability is confirmed to arise from the 
illuminated side of the inflated disc (bulge) while the optical modulation 
is produced by heating effects at the secondary star. \\

{\it 4)}~~A change in the relative proportion of rotational and orbital 
modulation amplitudes is found on a timescale of five years. These are
interpreted as a reduction in the dimensions of the accretion curtain 
accompained by an increase in the bulge size. \\

These observations indicate a long term
change in the accretion mode where FO\,Aqr has switched from a disc-fed
to a disk-overflow state triggered by changes in the mass accretion rate.

\begin{acknowledgements}

The authors wish to acknowledge the valuable help from ST\,ScI staff who
provided new calibration files for the square apertures of FOS instrument
before publication. J. Patterson is gratefully acknowledged for providing 
the new revised spin ephemeris quoted in the text. 
\end{acknowledgements}


\end{document}